# Protocol for pre-registration of a systematic review following the Preferred Reporting Items for Systematic Review and Meta-Analysis Protocols (PRISMA-P) guidelines by Shamseer et al. (2015)

**1. Working title:** Articulatory changes in speech following treatment for oral or oropharyngeal cancer: a systematic review

**2. Registration:** Registered with the International Prospective Register of Systematic Reviews (PROSPERO)

**3a Author contacts**

Thomas B. Tienkamp (TT), Centre for Language and Cognition Groningen (CLCG) Faculty of Arts, University of Groningen, Oude Kijk in 't Jatstraat 26, 9712EK Groningen, The Netherlands, t.b.tienkamp@rug.nl

Teja Rebernik (TR), Centre for Language and Cognition Groningen (CLCG) Faculty of Arts, University of Groningen, Oude Kijk in 't Jatstraat 26, 9712EK Groningen, The Netherlands, t.rebernik@rug.nl.

Defne Abur (DA), Centre for Language and Cognition Groningen (CLCG) Faculty of Arts, University of Groningen, Oude Kijk in 't Jatstraat 26, 9712EK Groningen, The Netherlands, d.abur@rug.nl

Rob, J.J.H. van Son (RvS), Netherlands Cancer Institute, Amsterdam, The Netherlands, Plesmanlaan 121, 1066 CX Amsterdam, The Netherlands. r.v.son@nki.nl

Sebastiaan A.H.J. de Visscher (SdV), Department of Oral and Maxillofacial Surgery, University Medical Centre Groningen, University of Groningen, Hanzeplein 1, 9713 GZ, Groningen, The Netherlands, s.a.h.j.de.visscher@umcg.nl

Max J.H. Witjes (MJHW), Department of Oral and Maxillofacial Surgery, University Medical Centre Groningen, University of Groningen, Hanzeplein 1, 9713 GZ, Groningen, The Netherlands, m.j.h.witjes@umcg.nl

Martijn B. Wieling (MBW), Centre for Language and Cognition Groningen (CLCG) Faculty of Arts, University of Groningen, Oude Kijk in 't Jatstraat 26, 9712EK Groningen, The Netherlands, m.b.wieling@rug.nl

Corresponding author: Thomas B. Tienkamp, t.b.tienkamp@rug.nl

## 3b Author contributions

In Table 1 below, the planned contribution of each author is provided following the CRediT author statement. Note that these are the planned contributions. Author contributions may be changed in the final report to correct for discrepancies in intended and actual contributions. Thomas Tienkamp (TT) will act as principal investigator.

**Table 1.** Author contributions during the systematic review.

| Task | TT | TR | DA | RvS | SdV | MJHW | MBW |
|---|---|---|---|---|---|---|---|
| Conceptualisation | X | X | X | X | X | X | X |
| Data curation | X | | | | | | |
| Formal analysis | X | | | | | | |
| Funding acquisition | | | | N/A | | | |
| Investigation | X | X | X | X | X | X | X |
| Methodology | X | X | X | X | X | X | X |
| Project administration | X | | | | | X | X |
| Resources | X | | | | | | X |
| Software | | | | N/A | | | |
| Supervision | | | | N/A | | | |
| Validation | | | | N/A | | | |
| Visualisation | X | | | | | | |
| Writing: original draft | X | | | | | | |
| Writing: review & editing | X | X | X | X | X | X | X |

**4. Amendments:** not applicable.

**5. Support**

The systematic review is part of the doctoral research from Thomas Tienkamp (TT), funded by the CLCG, University of Groningen. The University of Groningen also provides support in terms of database access and advice from the research librarian.

**Conflicts of interest**

# INTRODUCTION

## 6. Rationale

Oral cancer (OC) is an umbrella term for tumours spanning from the vermillion border of the lips, to the buccal mucosa, the front two thirds of the tongue, floor of mouth, the upper and lower gums, retromolar pad and hard palate (El-Naggar et al., 2017). Although in the latest update of the TNM staging system cancers of the vermillion border are considered skin cancers, they will be included in this systematic review since the treatment will affect oral function. Worldwide, oral cancer affects an estimated 355.000 people each year, comprising about 2% of all cancer incidences (Bray et al., 2018). Squamous cell carcinoma is the most common type of oral cancer, accounting for 9 out of 10 cases (Bagan et al., 2010). Once diagnosed, OC treatment depends on several factors such as the site and size of the tumour, the aetiology, and the preferences and clinical condition of the patients themselves. Generally, treatment consists of either surgical resection or a (chemo)radiation-based therapy (Constantinescu & Rieger, 2019). Treatment modalities may be combined, especially for larger tumours, as local recurrence remains high (Cohan et al., 2009). In those instances, the patient receives postoperative radiation therapy.

Among all cancer types, treatment of OC has one of the highest risks of loss or damage to important, or even vital, functions (Kreeft et al., 2009). For surgical treatment, depending on the location of the resection, the patient may experience problems with swallowing (dysphagia) (Borggreven et al., 2007; de Vicente et al., 2021; Lam & Samman, 2013) or altered/loss of sensation (Loewen et al., 2010). For radiation-based treatments, patients might experience problems with swallowing (Lazarus et al., 2007; Logemann et al., 2008), altered taste (Hovan et al., 2010), dry mouth (xerostomia; Chi et al., 2015), and an immobile jaw (trismus; Lee et al., 2015). All functional problems may contribute to a reduced quality of life post-treatment (Dwivedi et al., 2012; Epstein et al., 1999; Mowry et al., 2006).

Another functional issue that often arises, are problems with speech. Tissue loss or scar tissue as a result of surgery or limited tongue mobility due to radiation side-effects complicate articulation (Constantinescu & Rieger, 2019; Jacobi et al., 2013; Laaksonen et al., 2011). The resulting speech may be less intelligible, which complicates everyday communication, workplace reintegration, and could lead to social isolation and a reduced quality of life (Dwivedi et al., 2009; Epstein et al., 1999; Meyer et al., 2004). Moreover, patients rank speech in their top priorities post-treatment (Rogers et al., 2002; Tschiesner et al., 2013). Therefore, an understanding of the induced changes in speech is of paramount importance, as it can inform rehabilitation strategies and improve post-treatment quality of life.

The speech of OC patients has been analysed in a number of studies by means of perceptual and acoustic methods. Perceptual evaluations found that important factors affecting intelligibility are the size of the resection, such that better intelligibility was found after smaller excisions

(Bressmann et al., 2004; Nicoletti et al., 2004; Pauloski et al., 1998); if the tongue was more mobile (Bressmann et al., 2004; Matsui et al., 2007); and when the patient did not receive adjuvant radiation therapy (Matsui et al., 2007). Acoustic studies have analysed the speech signal in more detail and found that OC patients experience the most problems with fricatives, especially sibilants (Acher et al., 2014; Jacobi et al., 2013; Laaksonen et al., 2011; Zhou et al., 2011) and plosives (de Bruijn et al., 2009; Jacobi et al., 2013). Moreover, a reduction of the Vowel Space Area (VSA) may be observed (de Bruijn et al., 2009; Jacobi et al., 2013; Takatsu et al., 2017).

Even though perception and acoustic studies contribute greatly to the understanding of changes in speech, these studies provide indirect evidence of treatment induced articulatory patterns. In order to investigate the ontogenesis of the speech problems, the articulatory movements of the tongue, jaw, and lips need to be tracked directly. This has been done with a variety of methods, such as ultrasound (Bressmann et al., 2005, 2007, 2009, 2010); electropalatography (Fletcher, 1988); cine-MRI (Stone et al., 2014; Zhou et al., 2011, 2013); and real-time MRI (Hagedorn et al., 2021). The findings of these studies suggest that surgical treatment induces asymmetrical movement (Bressmann et al., 2005; Stone et al., 2014); reduced midsagittal tongue grooving (Bressmann et al., 2005, 2007); more backed pronunciation of sibilants (Zhou et al., 2011, 2013); and less complex vocal tract shaping (Hagedorn et al., 2021).

Given the quality-of-life impact of the treatment and the rated importance of speech by patients, a systematic review synthesising the changes in articulatory movements is important for two reasons. First, understanding which articulatory movements are impaired may inform speech language therapists in designing more effective rehabilitation, as current standardised therapies are almost non-existent (Bressmann, 2021). Second, the systematic review may be informative for surgeons, too, as treatment methods may be developed that reduce articulatory deficits.

*Previous reviews*

A scoping search of the literature indicates that no study up until now has systematically synthesised the literature on the articulatory changes following OC treatment. Still, some relevant reviews may be noted. For example, Balaguer et al. (2020) systematically reviewed the changes in the acoustic speech signal following treatment for oral or oropharyngeal cancer. Schuster and Stelzle (2012) reviewed the speech and speech-related outcome measures that have been used to characterise the speech of oral cancer patients. However, since their review covered intelligibility, acoustic, and articulatory methods, the authors could not provide much information regarding the experimental design or a detailed account of the results. Blyth et al. (2015) reviewed speech and swallowing following partial glossectomy, but solely focused on rehabilitation studies. Dwivedi et al. (2009) reviewed speech outcomes of oral and oropharyngeal cancer treatment between 2000 and 2008. Their review focused on questionnaire, perceptual, and acoustic evaluations, but did not include articulatory studies. Jacobi et al. (2010) reviewed the voice and speech outcomes after chemoradiation based treatment for head and neck cancer, and did not focus on surgically treated patients. Lastly, Lam & Samman (2013) reviewed speech and swallowing following tongue surgery and a radial forearm free flap reconstruction (RFFF). Although they included two articulatory studies in their review, many studies could not be included due to their inclusion criterion of a minimal sample size of 10. However, due to the technical and temporal demands of articulatory methods, sample sizes are generally small. For example, the cited papers by Bressmann and colleagues had an average sample size of four patients (range: 1-12).

We also searched for protocols of ongoing systematic reviews pertaining to our topic on PROSPERO. Two protocols were found: the already published review by Balaguer et al. (2020) and a protocol dating back to January 2019 on the effects of surgical treatment for oral and maxillofacial cancer on speech. This protocol intends to review the outcomes reported by both patient and a 'proxy' and thus diverts from the specific aims of our review as we intend to only review articulatory studies.

## 7. Objectives

The aim of this systematic review is to evaluate the effect of OC treatment on articulatory patterns in speech in terms of tongue, jaw, and lip movement. To this end, the proposed systematic review will answer the following questions:

*Main research question:* To what extent does treatment for oral or oropharyngeal cancer affect the articulatory movements of the tongue, jaw, and lips as compared to individuals without oral cancer?

The systematic review additionally aims to answer the following sub questions:
   (1) To what extent are articulatory deficits related to the tumour‑node‑metastasis staging of the tumour?
   (2) To what extent are articulatory deficits related to the place of the tumour?
   (3) To what extent are articulatory deficits related to treatment modality?
   (4) Is there a difference in the severity of the articulatory deficit found between patients with and without adjuvant radiation therapy?
   (5) How does the time post-treatment relate to the severity of the articulatory deficit?

Table 2 details the key components of our main research question using the PICO (Population, Intervention, Comparator, Outcome) tool (Schardt et al., 2007).

**Table 2.** Description of the key components of the main research question

|   | Component | Description |
|---|---|---|
| **P** | Population | Individuals (18+) treated for oral or oropharyngeal cancer |
| **I** | Intervention | Surgical or radiation-based treatment for oral or oropharyngeal cancer |
| **C** | Comparator | Individuals without oral cancer or earlier recordings (e.g., pre-treatment) of the same patient |
| **O** | Outcome | Articulatory movements of the tongue, jaw, or lips as captured by either ultrasound, magnetic resonance imaging, electropalatography, videofluoroscopy, electromagnetic articulography or X-ray |

**METHODS**

**8. Eligibility Criteria**

Studies will be selected according to the criteria outlined below.

(a) *Population*: A study needs to include one or more individuals (18+) who have received treatment for oral or oropharyngeal cancer. These types of cancer are defined as the existence of a tumour from the vermillion border of the lips, to the buccal mucosa, the front two thirds of the tongue, floor of mouth, the upper and lower gums, retromolar pad and hard palate. Tumours on the posterior one third of the tongue is included as well. Type of treatment (surgery or (chemo)radiation) does not constitute an exclusion criterion.

(b) *Relevant outcomes*: articulatory movements of the tongue, jaw, and lips (e.g., total displacement or tongue grooving); the experimental paradigm; patient characteristics (age and gender); treatment characteristics (surgery, reconstruction, (chemo)radiation, radiation dosage); and tumour characteristics (TNM staging, location).

(c) *Methods and data collection*: Studies need to assess articulatory movements directly, either through an experimental paradigm or other objective assessment. Studies that solely use questionnaire-based, perceptual, or acoustic evaluations will be excluded. Case studies and observational studies will be included as long as they report an objective and direct measure of articulatory movements.

(d) *Study design*: Either experimental or observational studies using the following designs: cross-sectional, longitudinal, case-control, and case studies (including case series).

The eligibility criteria outlined above were translated into questions that were meant to facilitate the screening process and provided in Table 3. The questions were trialled on several articles to assess their comprehensiveness, and have been adapted accordingly. The answer to all questions needs to be 'yes' in order for the study to be included in the systematic review.

**Table 3.** Screening questions used to assess the eligibility criteria for the systematic review.

| Component | Question | Yes | No |
|---|---|---|---|
| **Population** | Does the study include at least one individual who has been treated for oral cancer? | Include | Exclude |
| **Relevant outcomes** | Does the study assess the speech articulation of the participants objectively? | Include | Exclude |
| | Does the study describe patient, treatment or tumour characteristics? | Include | Exclude |
| **Method and data collection** | Does the study collect and analyse data pertaining to the articulatory movements of the participant? | Include | Exclude |
| **Study design** | Does the study report on an experimental or observational study using a cross-sectional, longitudinal, case-control, or case study design? | Include | Exclude |
| **Source** | Has the study been subjected to peer-review (either in a journal or full paper conference proceedings)? | Include | Exclude |

## 9. Information sources

Potentially eligible articles will be collected from several databases which we will search through according to our search strategy that will be outlined below. The exact search string per database can be found in **Appendix A**. We will use the following databases, which are either open-access or accessible through university subscriptions:

- PubMed
- Embase
- Scopus
- Web of Science
- PsycInfo

Additionally, the reference lists of related review articles (e.g., Balaguer et al., 2020; Schuster & Stelzle, 2012) will be hand-searched for missing but relevant studies.

Since the final search is expected to take place in August 2022, we set the 31st of July 2022 as the upper time limit for study inclusion. No lower limit was imposed since articles suitable for our systematic review are expected to be scarce. As our search terms were formulated in English (see Section 10), we expect most studies to be written in English. However, studies in Dutch, German and French, if detected by our search terms, will be included as well due to the authors' proficiency in these languages.

Corresponding authors will be contacted for inaccessible papers that passed the title and abstract screening (see Section 11). Authors will be contacted twice via email, with the second email sent one week after the first if no response was received. If no response is obtained after two weeks, data extraction will proceed without these items.

**10. Search strategy**

Based on previous systematic reviews and consultation with a research librarian, relevant search terms were selected (see Table 4). We refer to **Appendix A** for our complete queries. The first part of the query pertains to the population or intervention and is limited to the abstract and title of the studies. As we are interested in experimental or observational studies, we expect that the population and intervention are mentioned in the title or abstract.

**Table 4.** Keywords used in the systematic search.

| Query relating to | Keyword |
|---|---|
| Population and disease | Oral squamous cell carcinoma, squamous cell carcinoma, Oral cancer, Oral tumo*, Oral carcinoma, Mouth cancer, Mouth tumo*, Mouth carcinoma, Oropharyngeal cancer, Oropharyngeal tumo*, Oropharyngeal carcinoma, Head and neck cancer, Head and neck tumo*, Head and neck carcinoma, Facial cancer, Facial tumo*, Facial carcinoma, Tongue cancer, Tongue tumo*, Tongue carcinoma, Glossectom*, Post-glossectom*, Postglossectom* |
| Outcome | Movement, articulation, speech, intelligibility, acousti*, phoneti*, speech perception, speech therapy, tongue displacement, tongue motion, tongue positio*, lingual movement, lingual displacement, jaw displacement, tongue movement, jaw movement, lip displacement, lip movement, lip aperture, asymmetr*, symmetr*, concav*, tongue tip elevation |
| Method | Magnetic resonance imag*, MRI, rt-MRI, rtMRI, Real-time MRI, cine-MRI, ultrasound, UTI, ultrasound tongue imaging, EMA, electromagnetic articulography, EPG, electropalatography, palatography, vocal tract, linguopalatal contact, Videofluoroscop*, X-ray, X-ray microbeam |

The second part of the query relates to the speech outcomes following the intervention (i.e., the outcome). In order to only review studies that used articulatory methods, we included a third part to our query that describes the method. By using the boolean operator 'AND' for each part of our query, we eliminated the use of ultrasound or MRI to map the tumour for treatment rather than linguistic purposes.

We evaluated the comprehensiveness of the search strategy by performing trial searches in each of the chosen databases in two ways. First, we screened the first 50 results in each database to optimise our search string by looking for additional relevant keywords or strings. Second, we compared the search results to a hand searched list in order to ensure that our search string was able to capture these documents.

## 11. Study records

### 11a Data management

All articles identified by the literature as detailed above will be exported as .ris/.nbib files to the Zotero reference manager in an allocated folder for the systematic review. Primary reviewer (TT) will screen the folder for duplicates using the "Duplicate items' tab that is available in Zotero. Next, the title and abstract of each item will be imported into Rayyan, a free-access online programme that facilitates the screening process between multiple reviewers (Ouzzani et al., 2016). Both TT and TR will screen the articles for inclusion following the process detailed in Section 11b. After the initial screening, full texts of the eligible articles will be imported into Zotero for the next step of the selection process. Relevant data will be extracted by TT from the eligible articles that passed the full-text stage using Excel for the following categories: (1) identification; (2) general study characteristics; (3) participant characteristics; (4) tumour characteristics; (5) treatment characteristics; (6) experiment information; (7) quantitative results; (8) qualitative results (see **Appendix B** for a full overview).

### 11b Selection process

The primary reviewers (TT and TR) will screen the title and abstract of each study that has been imported into Rayyan after deduplication. The screening questions (Table 3) that are based on the eligibility criteria will facilitate the screening process. Then, the full reports for all titles that passed the initial screening, or for which there was uncertainty, will be obtained and imported into Rayyan by TT. TT and TR will screen these studies for final inclusion using the same screening questions. Any uncertainty, both in the abstract-screening and full-text stage, will be resolved through discussion between the reviewers. In the case the reviewers still disagree, a third reviewer (MBW) will be consulted to resolve the uncertainty. Reason for exclusion in the full-article stage will be recorded. The final list of included studies will be imported into Excel and the full reports into Zotero.

**11c Data collection process**

The information from the included studies will be collected using data extraction forms (see **Appendix B** for the full form). These forms were piloted on several studies prior to protocol registration. Data extraction will be conducted by the first reviewer (TT) in Excel. In case of uncertainty, the second reviewer (TR) will be consulted.

**12. Data items**

From the included studies, the following categories of information will be extracted:

- Identification (e.g., authors, year of publication, and a full reference)
- General study characteristics (e.g., study design, language, and location)
- Participant characteristics (e.g., sample size, age, gender, and language)
- Tumour characteristics (e.g., TNM staging, location)
- Treatment characteristics (e.g., procedure type (surgical or radiation-based) and its accompanying details (e.g., reconstruction or radiation dosage)).
- Experiment information (e.g., procedure, method, goal)
- Quantitative results (e.g., modelling method, descriptive results, and significance level)
- Qualitative results (e.g., any descriptive results regarding the articulatory patterns that are elaborated on in the paper)
- Individual data (if existent)

See **Appendix B** for the entire data extraction form as well as a filled-out example based on the study by Bressmann et al. (2007). If individual patient data are elaborated on by the study authors, these will be included as well, paired with the characteristics of the patient (e.g., age, TNM stage, tumour location).

**13. Outcomes and prioritisation**

Primary outcome of interest:
The main outcomes of interest are the articulatory patterns of individuals who have received treatment for oral or oropharyngeal cancer. The articulatory patterns will be assessed on the basis of experimental or observational studies using articulatory methods. Example outcome variables for lingual movement are: tongue velocity, concavity, asymmetry, and tongue motion patterns in terms of anterior-posterior and superior-inferior movement. Variables concerning lip and jaw movement are of interest as well (e.g., peak velocity, lip aperture, etc.). Studies that included only a patient group (e.g., case series or cross-sectional) as well as studies that included both a patient and control group (e.g., case-control studies) will be included.

Additional (secondary) outcomes of interest:
1. The effects of disease stage (i.e., TNM staging) on articulatory patterns, as some perceptual studies have found a link between the size of the resection and speech outcomes (e.g., Bressmann et al., 2004; Nicoletti et al., 2004).
2. The effect of the location of the tumour on the articulatory patterns. Different parts of the tongue are used for different sounds (e.g., the tongue tip is used to produce /t/, but not /k/). We therefore expect that speech outcomes are modulated by the location of the tumour. One subdivision that is often made is between tumours on the anterior part of the tongue versus those on the base of the tongue (e.g., Takatsu et al., 2017). Another distinction that is often made is whether a floor of the mouth resection was necessary or not (e.g., Laaksonen et al., 2011).
3. The effect of treatment modality on the articulatory patterns. On the anatomical level, one might predict differences between treatment modalities (surgical versus radiation-based treatment) with regards to articulatory deficits as radiation-based treatment preserves the organ. However, Jacobi et al. (2013) showed that patients treated with chemoradiation still experienced speech problems on the acoustic level. Whether there are differences in terms of articulatory movements, and hence the nature of the articulatory deficits, is not known.
4. The impact of adjuvant radiation therapy on articulatory patterns. Radiation therapy might induce an additional stiffening of the tongue and jaw which could impact their movement in speech production. For example, individuals with adjuvant radiation therapy had worse speech outcomes compared to those without (Nicoletti et al., 2004).
5. In the case of longitudinal studies, the temporal development of articulatory changes may be analysed in order to establish whether changes in articulatory movements are lasting. Some studies have found that acoustic values return to control-like levels (Zhou et al., 2011) whereas others report on significant differences after one year (Laaksonen et al., 2011).
6. If enough participant details are provided in the studies, characteristics such as age and gender might be compared.

## 14. Risk of bias in individual studies

To assess the risk of bias within included studies, the standardised critical appraisal checklist for non-randomised experimental studies (quasi-experimental) as provided by the Joanna Briggs Institute (JBI) will be used (Tufanaru et al., 2020). The checklist assesses the methodological quality of an individual study on the following domains using nine yes/no questions: (1) chosen variables; (2) chosen participant groups; (3) outcome measurement (reliability); and (4) statistical analysis. Studies of poor quality will be excluded from data synthesis, but explicitly addressed and discussed in the narrative synthesis. The quality of each study will be assessed by the primary reviewer (TT). To assess risk of bias for case series, the critical appraisal checklist for case series by the JBI will be used (Munn et al., 2020). Any uncertainty will be resolved through consultation with the second reviewer (TR).

## 15. Data synthesis

The primary synthesis methods outlined by Moher et al. (2015) assume a form of quantitative synthesis (15a-c). However, our planned systematic review is not suitable for quantitative synthesis. Due to the considerable variation in the included articles in terms of experimental paradigms and stimuli, we do not intend to perform a meta-analysis. Instead, we aim to provide a narrative synthesis (see details below). Should enough individual data be provided by the studies, a meta-analysis will be performed on patient characteristics.

## 15d. Narrative synthesis

The present systematic review will perform a narrative synthesis of extracted group data which will summarise and explain the characteristics and outcomes of the included studies. Outcomes will be reported following the Synthesis Without Meta-Analysis (SWiM) guideline (Campbell et al., 2020). The SWiM guideline is intended for systematic reviews of quantitative effects for which a meta-analysis is not possible due to the diversity in stimuli or experimental paradigms of the included studies. The guideline provides a nine-item checklist that promotes transparency for narrative syntheses and is meant to be used as an extension of the PRISMA guideline (Campbell et al., 2020; Moher et al., 2015). Visuals in the form of tables describing the characteristics of included studies will be included. The review will be divided into subsections per sub-question in order to answer the main review question (e.g., a section comparing and contrasting the results for different tumour locations).

## 16. Metabias

We will compare the published study to a pre-registered protocol of the study, if available. If this is not available, the methods and results section will be compared in order to detect outcome reporting bias (Chan & Altman, 2005). Furthermore, the publication bias will be assessed by looking at the proportion of studies that found significant differences between OC patients and controls versus studies that found no differences. This will serve as a proxy of the likelihood of journals publishing null results.

## 17. Confidence in cumulative evidence

The quality of the body of evidence will be assessed for each outcome using the GRADE guidelines (Grades of Recommendation, Assessment, Development and Evaluation) from the

GRADE working group (Guyatt et al., 2008). Since our systematic will most likely not include randomised controlled trials, the quality of the body of evidence will start at a low rating.

**Transparency**

This protocol was pre-registered when scoping searches were completed. None of the review authors is an author of one of the studies that are likely to be included in the systematic review.

# REFERENCES

Acher, A., Perrier, P., Savariaux, C., & Fougeron, C. (2014). Speech production after glossectomy: Methodological aspects. *Clinical Linguistics & Phonetics*, *28*(4), 241–256. https://doi.org/10.3109/02699206.2013.802015

Balaguer, M., Pommée, T., Farinas, J., Pinquier, J., Woisard, V., & Speyer, R. (2020). Effects of oral and oropharyngeal cancer on speech intelligibility using acoustic analysis: Systematic review. *Head & Neck*, *42*(1), 111–130. https://doi.org/10.1002/hed.25949

Bagan, J., Sarrion, G., & Jimenez, Y. (2010). Oral cancer: clinical features. Oral oncology, 46(6), 414-417. https://doi.org/10.1016/j.oraloncology.2010.03.009

Blyth, K. M., McCabe, P., Madill, C., & Ballard, K. J. (2015). Speech and swallow rehabilitation following partial glossectomy: A systematic review. *International Journal of Speech-Language Pathology*, *17*(4), 401–410. https://doi.org/10.3109/17549507.2014.979880

Borggreven, P. A., Verdonck-de Leeuw, I., Rinkel, R. N., Langendijk, J. A., Roos, J. C., David, E. F. L., de Bree, R., & Leemans, C. R. (2007). Swallowing after major surgery of the oral cavity or oropharynx: A prospective and longitudinal assessment of patients treated by microvascular soft tissue reconstruction. *Head & Neck*, *29*(7), 638–647. https://doi.org/10.1002/hed.20582

Bray, F., Ferlay, J., Soerjomataram, I., Siegel, R. L., Torre, L. A., & Jemal, A. (2018). Global cancer statistics 2018: GLOBOCAN estimates of incidence and mortality worldwide for 36 cancers in 185 countries. *CA: A Cancer Journal for Clinicians*, *68*(6), 394–424. https://doi.org/10.3322/caac.21492

Bressmann, T. (2021). Speech Disorders Related to Head and Neck Cancer: Laryngectomy, Glossectomy, and Velopharyngeal and Maxillofacial Defects. In J. S. Damico, N. Müller, & M. J. Ball (Eds.), *The Handbook of Language and Speech Disorders* (2nd ed., pp. 495–527). Wiley-Blackwell.

Bressmann, T., Ackloo, E., Heng, C.-L., & Irish, J. C. (2007). Quantitative Three-Dimensional Ultrasound Imaging of Partially Resected Tongues. *Otolaryngology–Head and Neck Surgery*, *136*(5), 799–805. https://doi.org/10.1016/j.otohns.2006.11.022

Bressmann, T., Flowers, H., Wong, W., & Irish, J. C. (2010). Coronal view ultrasound imaging of movement in different segments of the tongue during paced recital: Findings from four normal speakers and a speaker with partial glossectomy. *Clinical Linguistics & Phonetics*, *24*(8), 589–601. https://doi.org/10.3109/02699201003687309

Bressmann, T., Rastadmehr, O., Smyth, R., & Irish, J. C. (2009). Ultrasound imaging of tongue movement and speech in partial glossectomy patients. *Oral Oncology Supplement*, *3*(1), 188. https://doi.org/10.1016/j.oos.2009.06.487

Bressmann, T., Sader, R., Whitehill, T. L., & Samman, N. (2004). Consonant intelligibility and tongue motility in patients with partial glossectomy. *Journal of Oral and Maxillofacial Surgery*, *62*(3), 298–303. https://doi.org/10.1016/j.joms.2003.04.017

Bressmann, T., Thind, P., Uy, C., Bollig, C., Gilbert, R. W., & Irish, J. C. (2005). Quantitative three‑dimensional ultrasound analysis of tongue protrusion, grooving and symmetry: Data from 12 normal speakers and a partial glossectomee. *Clinical Linguistics & Phonetics*, *19*(6–7), 573–588. https://doi.org/10.1080/02699200500113947

Campbell, M., McKenzie, J. E., Sowden, A., Katikireddi, S. V., Brennan, S. E., Ellis, S., Hartmann-Boyce, J., Ryan, R., Shepperd, S., Thomas, J., Welch, V., & Thomson, H. (2020). Synthesis without meta-analysis (SWiM) in systematic reviews: Reporting guideline. *BMJ*, l6890. https://doi.org/10.1136/bmj.l6890

Chan, A.-W., & Altman, D. G. (2005). Identifying outcome reporting bias in randomised trials on PubMed: Review of publications and survey of authors. *BMJ*, *330*(7494), 753. https://doi.org/10.1136/bmj.38356.424606.8F

Chi, A. C., Day, T. A., & Neville, B. W. (2015). Oral cavity and oropharyngeal squamous cell carcinoma—An update. *CA: A Cancer Journal for Clinicians*, *65*(5), 401–421. https://doi.org/10.3322/caac.21293

Cohan, D. M., Popat, S., Kaplan, S. E., Rigual, N., Loree, T., & Hicks, W. L. (2009). Oropharyngeal cancer: Current understanding and management. *Current Opinion in Otolaryngology & Head & Neck Surgery*, *17*(2), 88–94. https://doi.org/10.1097/MOO.0b013e32832984c0

Constantinescu, G., & Rieger, J. M. (2019). Speech Deficits Associated with Oral and Oropharyngeal Carcinomas. In P. C. Doyle (Ed.), *Clinical Care and Rehabilitation in Head and Neck Cancer* (pp. 265–279). Springer International Publishing. https://doi.org/10.1007/978-3-030-04702-3_16

de Bruijn, M. J., ten Bosch, L., Kuik, D. J., Quené, H., Langendijk, J. A., Leemans, C. R., & Verdonck-de Leeuw, I. M. (2009). Objective Acoustic-Phonetic Speech Analysis in Patients Treated for Oral or Oropharyngeal Cancer. *Folia Phoniatrica et Logopaedica*, *61*(3), 180–187. https://doi.org/10.1159/000219953

de Vicente, J. C., Rúa-Gonzálvez, L., Barroso, J. M., Valle-Fernández, Á. F. del, Villalaín, L. de, Peña, I., & Cobo, J. L. (2021). Functional results of swallowing and aspiration after oral cancer treatment and microvascular free flap reconstruction: A retrospective observational assessment. *Journal of Cranio-Maxillofacial Surgery*, *49*(10), 959–970. https://doi.org/10.1016/j.jcms.2021.04.015

Dwivedi, R. C., Kazi, R. A., Agrawal, N., Nutting, C. M., Clarke, P. M., Kerawala, C. J., Rhys-Evans, P. H., & Harrington, K. J. (2009). Evaluation of speech outcomes following treatment of oral and oropharyngeal cancers. *Cancer Treatment Reviews*, *35*(5), 417–424. https://doi.org/10.1016/j.ctrv.2009.04.013

Dwivedi, R. C., St.Rose, S., Chisholm, E. J., Youssefi, P., Hassan, M. S. U., Khan, A. S., Elmiyeh, B., Kerawala, C. J., Clarke, P. M., Nutting, C. M., Rhys-Evans, P. H., Harrington, K. J., & Kazi, R. (2012). Evaluation of factors affecting post-treatment quality of life in oral and oropharyngeal cancer patients primarily treated with curative surgery: An exploratory study. *European Archives of Oto-Rhino-Laryngology*, *269*(2),

591–599. https://doi.org/10.1007/s00405-011-1621-z

El-Naggar, A. K., Chan, J. K. C., Grandis, J. R., Takata, T., & Slootweg, P. J. (Eds.). (2017). *WHO classification of head and neck tumours* (4th ed). International agency for research on cancer.

Epstein, J. B., Emerton, S., Kolbinson, D. A., Le, N. D., Phillips, N., Stevenson-Moore, P., & Osoba, D. (1999). Quality of life and oral function following radiotherapy for head and neck cancer. *Head & Neck: Journal for the Sciences and Specialties of the Head and Neck*, *21*(1), 1–11.

Fletcher, S. G. (1988). Speech Production Following Partial Glossectomy. *Journal of Speech and Hearing Disorders*, *53*(3), 232–238. https://doi.org/10.1044/jshd.5303.232

Guyatt, G. H., Oxman, A. D., Vist, G. E., Kunz, R., Falck-Ytter, Y., Alonso-Coello, P., & Schünemann, H. J. (2008). GRADE: An emerging consensus on rating quality of evidence and strength of recommendations. *BMJ*, *336*(7650), 924–926. https://doi.org/10.1136/bmj.39489.470347.AD

Hagedorn, C., Kim, J., Sinha, U., Goldstein, L., & Narayanan, S. S. (2021). Complexity of vocal tract shaping in glossectomy patients and typical speakers: A principal component analysis. *The Journal of the Acoustical Society of America*, *149*(6), 4437–4449. https://doi.org/10.1121/10.0004789

Hovan, A. J., Williams, P. M., Stevenson-Moore, P., Wahlin, Y. B., Ohrn, K. E. O., Elting, L. S., Spijkervet, F. K. L., Brennan, M. T., & Dysgeusia Section, Oral Care Study Group, Multinational Association of Supportive Care in Cancer (MASCC)/International Society of Oral Oncology (ISOO). (2010). A systematic review of dysgeusia induced by cancer therapies. *Supportive Care in Cancer*, *18*(8), 1081–1087. https://doi.org/10.1007/s00520-010-0902-1

Jacobi, I., van der Molen, L., Huiskens, H., van Rossum, M. A., & Hilgers, F. J. M. (2010). Voice and speech outcomes of chemoradiation for advanced head and neck cancer: A systematic review. *European Archives of Oto-Rhino-Laryngology*, *267*(10), 1495–1505. https://doi.org/10.1007/s00405-010-1316-x

Jacobi, I., van Rossum, M. A., van der Molen, L., Hilgers, F. J. M., & van den Brekel, M. W. M. (2013). Acoustic Analysis of Changes in Articulation Proficiency in Patients with Advanced Head and Neck Cancer Treated with Chemoradiotherapy. *Annals of Otology, Rhinology & Laryngology*, *122*(12), 754–762. https://doi.org/10.1177/000348941312201205

Kreeft, A. M., van der Molen, L., Hilgers, F. J., & Balm, A. J. (2009). Speech and swallowing after surgical treatment of advanced oral and oropharyngeal carcinoma: A systematic review of the literature. *European Archives of Oto-Rhino-Laryngology*, *266*(11), 1687–1698. https://doi.org/10.1007/s00405-009-1089-2

Laaksonen, J.-P., Rieger, J., Harris, J., & Seikaly, H. (2011). A longitudinal acoustic study of the effects of the radial forearm free flap reconstruction on sibilants produced by tongue cancer patients. *Clinical Linguistics & Phonetics*, *25*(4), 253–264.

https://doi.org/10.3109/02699206.2010.525681

Lam, L., & Samman, N. (2013). Speech and swallowing following tongue cancer surgery and free flap reconstruction – A systematic review. *Oral Oncology*, *49*(6), 507–524. https://doi.org/10.1016/j.oraloncology.2013.03.001

Lazarus, C., Logemann, J. A., Pauloski, B. R., Rademaker, A. W., Helenowski, I. B., Vonesh, E. F., MacCracken, E., Mittal, B. B., Vokes, E. E., & Haraf, D. J. (2007). Effects of radiotherapy with or without chemotherapy on tongue strength and swallowing in patients with oral cancer. *Head & Neck: Journal for the Sciences and Specialties of the Head and Neck*, *29*(7), 632–637. https://doi.org/10.1002/hed.20577

Lee, L.-Y., Chen, S.-C., Chen, W.-C., Huang, B.-S., & Lin, C.-Y. (2015). Postradiation trismus and its impact on quality of life in patients with head and neck cancer. *Oral Surgery, Oral Medicine, Oral Pathology and Oral Radiology*, *119*(2), 187–195. https://doi.org/10.1016/j.oooo.2014.10.003

Loewen, I. J., Boliek, C. A., Harris, J., Seikaly, H., & Rieger, J. M. (2010). Oral sensation and function: A comparison of patients with innervated radial forearm free flap reconstruction to healthy matched controls. *Head & Neck*, *32*(1), 85–95. https://doi.org/10.1002/hed.21155

Logemann, J. A., Pauloski, B. R., Rademaker, A. W., Lazarus, C. L., Gaziano, J., Stachowiak, L., Newman, L., MacCracken, E., Santa, D., & Mittal, B. (2008). Swallowing disorders in the first year after radiation and chemoradiation. *Head & Neck*, *30*(2), 148–158. https://doi.org/10.1002/hed.20672

Matsui, Y., Ohno, K., Yamashita, Y., & Takahashi, K. (2007). Factors influencing postoperative speech function of tongue cancer patients following reconstruction with fasciocutaneous/myocutaneous flaps—A multicenter study. *International Journal of Oral and Maxillofacial Surgery*, *36*(7), 601–609. https://doi.org/10.1016/j.ijom.2007.01.014

Meyer, T. K., Kuhn, J. C., Campbell, B. H., Marbella, A. M., Myers, K. B., & Layde, P. M. (2004). Speech Intelligibility and Quality of Life in Head and Neck Cancer Survivors: *The Laryngoscope*, *114*(11), 1977–1981. https://doi.org/10.1097/01.mlg.0000147932.36885.9e

Moher, D., Shamseer, L., Clarke, M., Ghersi, D., Liberati, A., Petticrew, M., Shekelle, P., Stewart, L. A., & PRISMA-P Group. (2015). Preferred reporting items for systematic review and meta-analysis protocols (PRISMA-P) 2015 statement. *Systematic Reviews*, *4*(1), 1. https://doi.org/10.1186/2046-4053-4-1

Mowry, S. E., Ho, A., LoTempio, M. M., Sadeghi, A., Blackwell, K. E., & Wang, M. B. (2006). Quality of Life in Advanced Oropharyngeal Carcinoma After Chemoradiation Versus Surgery and Radiation: *The Laryngoscope*, *116*(9), 1589–1593. https://doi.org/10.1097/01.mlg.0000233244.18901.44

Munn, Z., Barker, T. H., Moola, S., Tufanaru, C., Stern, C., McArthur, A., Stephenson, M., & Aromataris, E. (2020). Methodological quality of case series studies: An introduction to the JBI critical appraisal tool. *JBI Evidence Synthesis*, *18*(10), 2127–2133. https://doi.org/10.11124/JBISRIR-D-19-00099

Nicoletti, G., Soutar, D. S., Jackson, M. S., Wrench, A. A., Robertson, G., & Robertson, C. (2004). Objective Assessment of Speech after Surgical Treatment for Oral Cancer: Experience from 196 Selected Cases: *Plastic and Reconstructive Surgery*, *113*(1), 114–125. https://doi.org/10.1097/01.PRS.0000095937.45812.84

Ouzzani, M., Hammady, H., Fedorowicz, Z., & Elmagarmid, A. (2016). Rayyan—A web and mobile app for systematic reviews. *Systematic Reviews*, *5*(1), 210. https://doi.org/10.1186/s13643-016-0384-4

Pauloski, B. R., Logemann, J. A., Colangelo, L. A., Rademaker, A. W., McConnel, F. M., Heiser, M. A., Cardinale, S., Shedd, D., Stein, D., Beery, Q., & others. (1998). Surgical variables affecting speech in treated patients with oral and oropharyngeal cancer. *The Laryngoscope*, *108*(6), 908–916.

Rogers, S. N., Laher, S. H., Overend, L., & Lowe, D. (2002). Importance-rating using the University of Washington Quality of Life questionnaire in patients treated by primary surgery for oral and oro-pharyngeal cancer. *Journal of Cranio-Maxillofacial Surgery*, *30*(2), 125–132. https://doi.org/10.1054/jcms.2001.0273

Schardt, C., Adams, M. B., Owens, T., Keitz, S., & Fontelo, P. (2007). Utilization of the PICO framework to improve searching PubMed for clinical questions. *BMC Medical Informatics and Decision Making*, *7*(1), 16. https://doi.org/10.1186/1472-6947-7-16

Schuster, M., & Stelzle, F. (2012). Outcome measurements after oral cancer treatment: Speech and speech-related aspects—an overview. *Oral and Maxillofacial Surgery*, *16*(3), 291–298. https://doi.org/10.1007/s10006-012-0340-y

Shamseer, L., Moher, D., Clarke, M., Ghersi, D., Liberati, A., Petticrew, M., Shekelle, P., Stewart, L. A., & the PRISMA-P Group. (2015). Preferred reporting items for systematic review and meta-analysis protocols (PRISMA-P) 2015: Elaboration and explanation. *BMJ*, *349*(jan02 1), g7647–g7647. https://doi.org/10.1136/bmj.g7647

Stone, M., Langguth, J. M., Woo, J., Chen, H., & Prince, J. L. (2014). Tongue Motion Patterns in Post-Glossectomy and Typical Speakers: A Principal Components Analysis. *Journal of Speech, Language, and Hearing Research*, *57*(3), 707–717. https://doi.org/10.1044/1092-4388(2013/13-0085)

Takatsu, J., Hanai, N., Suzuki, H., Yoshida, M., Tanaka, Y., Tanaka, S., Hasegawa, Y., & Yamamoto, M. (2017). Phonologic and Acoustic Analysis of Speech Following Glossectomy and the Effect of Rehabilitation on Speech Outcomes. *Journal of Oral and Maxillofacial Surgery*, *75*(7), 1530–1541. https://doi.org/10.1016/j.joms.2016.12.004

Tschiesner, U., Sabariego, C., Linseisen, E., Becker, S., Stier-Jarmer, M., Cieza, A., & Harreus, U. (2013). Priorities of head and neck cancer patients: A patient survey based on the brief ICF core set for HNC. *European Archives of Oto-Rhino-Laryngology*, *270*(12), 3133–3142. https://doi.org/10.1007/s00405-013-2446-8

Tufanaru, C., Munn, Z., Aromataris, E., Campbell, J., & Hopp, L. (2020). Chapter 3: Systematic reviews of effectiveness. In E. Aromataris & Z. Munn (Eds.), *JBI Manual for Evidence Synthesis*. JBI. https://synthesismanual.jbi.global

Zhou, X., Stone, M., & Espy-Wilson, C. Y. (2011). A comparative acoustic study on speech of glossectomy patients and normal subjects. *Proceedings of Interspeech 2011: 12th Annual Conference of the International Speech Communication Association*, 517–520.

Zhou, X., Woo, J., Stone, M., & Espy-Wilson, C. (2013). A cine MRI-based study of sibilant fricatives production in post-glossectomy speakers. *2013 IEEE International Conference on Acoustics, Speech and Signal Processing*, 7780–7784. https://doi.org/10.1109/ICASSP.2013.6639178

**Appendix A:** Search strings per database

| Database | access | Search string | clarification |
|---|---|---|---|
| PubMed | https://pubmed.ncbi.nlm.nih.gov/ | (<br>Mouth Neoplasms[Mesh] OR<br>Oropharyngeal Neoplasms[Mesh] OR<br>Facial Neoplasms[Mesh] OR<br>Head and Neck Neoplasms[Mesh] OR<br>Tongue Neoplasms[Mesh] OR<br>Oral squamous cell carcinoma[Title/Abstract] OR<br>Squamous cell carcinoma[Title/Abstract] OR<br>Oral cancer[Title/Abstract] OR<br>Oral tumo*[Title/Abstract] OR<br>Oral carcinoma[Title/Abstract] OR<br>Mouth cancer[Title/Abstract] OR<br>Mouth tumo*[Title/Abstract] OR<br>Mouth carcinoma[Title/Abstract] OR<br>Oropharyngeal cancer[Title/Abstract] OR<br>Oropharyngeal tumo*[Title/Abstract] OR<br>Oropharyngeal carcinoma[Title/Abstract] OR<br>Head and neck cancer[Title/Abstract] OR<br>Head and neck tumo*[Title/Abstract] OR<br>Head and neck carcinoma[Title/Abstract] OR<br>Facial cancer[Title/Abstract] OR<br>Facial tumo*[Title/Abstract] OR<br>Facial carcinoma[Title/Abstract] OR<br>Tongue cancer[Title/Abstract] OR<br>Tongue tumo*[Title/Abstract] OR<br>Tongue carcinoma[Title/Abstract] OR<br>Glossectom*[Title/Abstract] OR<br>Post-glossectom*[Title/Abstract] OR<br>Postglossectom*[Title/Abstract]<br>)<br>AND<br>(<br>Articulation Disorders[Mesh] OR<br>Speech Intelligibility[Mesh] OR<br>movement[Title/Abstract] OR<br>articulation[Title/Abstract] OR<br>speech[Title/Abstract] OR<br>intelligibility[Title/Abstract] OR<br>voice[Title/Abstract] OR<br>acousti*[Title/Abstract] OR<br>phoneti*[Title/Abstract] OR<br>Speech perception[Title/Abstract] OR<br>Speech therapy[Title/Abstract] OR<br>tongue displacement[Title/Abstract] OR<br>tongue motion[Title/Abstract] OR<br>tongue positio*[Title/Abstract] OR<br>Lingual displacement[Title/Abstract] OR<br>Jaw displacement[Title/Abstract] OR<br>lingual movement[Title/Abstract] OR<br>Tongue movement[Title/Abstract] OR | [Title/Abstract] = limit search to title and abstract |
| | | Jaw movement[Title/Abstract] OR<br>lip displacement[Title/Abstract] OR<br>lip movement[Title/Abstract] OR<br>lip aperture[Title/Abstract] OR<br>asymmetr*[Title/Abstract] OR<br>symmetr*[Title/Abstract] OR<br>concav*[Title/Abstract] OR<br>tongue tip elevation[Title/Abstract]<br>)<br>AND<br>(<br>magnetic resonance imag*[Title/Abstract] OR<br>MRI[Title/Abstract] OR<br>rt-MRI[Title/Abstract] OR<br>rtMRI[Title/Abstract] OR<br>Real-time MRI[Title/Abstract] OR<br>cine-MRI[Title/Abstract] OR<br>ultrasound[Title/Abstract] OR<br>UTI[Title/Abstract] OR<br>ultrasound tongue imaging[Title/Abstract] OR<br>EMA[Title/Abstract] OR<br>Electromagnetic articulography[Title/Abstract] OR<br>EPG[Title/Abstract] OR<br>Electropalatography[Title/Abstract] OR<br>Palatography[Title/Abstract] OR<br>vocal tract[Title/Abstract] OR<br>linguopalatal contact[Title/Abstract] OR<br>Videofluoroscop*[Title/Abstract] OR<br>X-ray[Title/Abstract] OR<br>X-ray microbeam[Title/Abstract]<br>) | |
| PsychInfo | via EbscoHost | AB(<br>"Oral squamous cell carcinoma" OR<br>"squamous cell carcinoma" OR<br>"Oral cancer" OR<br>"Oral tumo*" OR<br>"Oral carcinoma" OR<br>"Mouth cancer" OR<br>"Mouth tumo*" OR<br>"Mouth carcinoma" OR<br>"Oropharyngeal cancer" OR<br>"Oropharyngeal tumo*" OR<br>"Oropharyngeal carcinoma" OR<br>"Head and neck cancer" OR<br>"Head and neck tumo*" OR<br>"Head and neck carcinoma" OR<br>"Facial cancer" OR<br>"Facial tumo*" OR<br>"Facial carcinoma" OR<br>"Tongue cancer" OR<br>"Tongue tumo*" OR<br>"Tongue carcinoma" OR<br>Glossectom* OR<br>Post-glossectom* OR | AB =<br>limit search to abstract |

| | | | Postglossectom*<br>)<br><br>AB(<br>"movement" OR<br>"articulation" OR<br>"speech" OR<br>"intelligibility" OR<br>"voice" OR<br>"acousti*" OR<br>"phoneti*" OR<br>"speech perception" OR<br>"Speech therapy" OR<br>"tongue displacement" OR<br>"Tongue motion" OR<br>"Tongue positio*" OR<br>"lingual movement" OR<br>"Lingual displacement" OR<br>"Jaw displacement" OR<br>"Tongue movement" OR<br>"Jaw movement" OR<br>"asymmetr*" OR<br>"symmetr*" OR<br>"Lip displacement" OR<br>"Lip movement" OR<br>"Lip aperture" OR<br>"concav*" OR<br>"tongue tip elevation"<br>)<br><br><br>AB(<br>"magnetic resonance imag*" OR<br>"MRI" OR<br>"rt-MRI" OR<br>"rtMRI" OR<br>"Real-time MRI" OR<br>"cine-MRI" OR<br>"ultrasound" OR<br>"UTI" OR<br>"ultrasound tongue imaging" OR<br>"EMA" OR<br>"Electromagnetic articulography" OR<br>"EPG" OR<br>"Electropalatography" OR<br>"Palatography" OR<br>"vocal tract" OR<br>"linguopalatal contact" OR<br>"Videofluoroscop*" OR<br>"X-ray" OR<br>"X-ray microbeam"<br>) | |
| Scopus | https://www.scopus.com/ | TITLE-ABS-KEY( | TITLE-ABS-KEY |

| | | "Oral squamous cell carcinoma" OR<br>"squamous cell carcinoma" OR<br>"Oral cancer" OR<br>"Oral tumo*" OR<br>"Oral carcinoma" OR<br>"Mouth cancer" OR<br>"Mouth tumo*" OR<br>"Mouth carcinoma" OR<br>"Oropharyngeal cancer" OR<br>"Oropharyngeal tumo*" OR<br>"Oropharyngeal carcinoma" OR<br>"Head and neck cancer" OR<br>"Head and neck tumo*" OR<br>"Head and neck carcinoma" OR<br>"Facial cancer" OR<br>"Facial tumo*" OR<br>"Facial carcinoma" OR<br>"Tongue cancer" OR<br>"Tongue tumo*" OR<br>"Tongue carcinoma" OR<br>Glossectom* OR<br>Post-glossectom* OR<br>Postglossectom*<br>)<br><br>TITLE-ABS-KEY(<br>"movement" OR<br>"articulation" OR<br>"speech" OR<br>"intelligibility" OR<br>"voice" OR<br>"acousti*" OR<br>"phoneti*" OR<br>"speech perception" OR<br>"Speech therapy" OR<br>"tongue displacement" OR<br>"Tongue motion" OR<br>"Tongue positio*" OR<br>"lingual movement" OR<br>"Lingual displacement" OR<br>"Jaw displacement" OR<br>"Tongue movement" OR<br>"Jaw movement" OR<br>"asymmetr*" OR<br>"symmetr*" OR<br>"Lip displacement" OR<br>"Lip movement" OR<br>"Lip aperture" OR<br>"concav*" OR<br>"tongue tip elevation"<br>)<br><br><br>TITLE-ABS-KEY( | =<br>limit search to title, abstract and keywords |
| | | "magnetic resonance imag*" OR<br>"MRI" OR<br>"rt-MRI" OR<br>"rtMRI" OR<br>"Real-time MRI" OR<br>"cine-MRI" OR<br>"ultrasound" OR<br>"UTI" OR<br>"ultrasound tongue imaging" OR<br>"EMA" OR<br>"Electromagnetic articulography" OR<br>"EPG" OR<br>"Electropalatography" OR<br>"Palatography" OR<br>"vocal tract" OR<br>"linguopalatal contact" OR<br>"Videofluoroscop*" OR<br>"X-ray" OR<br>"X-ray microbeam"<br>) | |
| Web of Science | https://www.webofscience.com/wos/woscc/advanced-search | ((TS=<br>("Oral squamous cell carcinoma" OR<br>"squamous cell carcinoma" OR<br>"Oral cancer" OR<br>"Oral tumo*" OR<br>"Oral carcinoma" OR<br>"Mouth cancer" OR<br>"Mouth tumo*" OR<br>"Mouth carcinoma" OR<br>"Oropharyngeal cancer" OR<br>"Oropharyngeal tumo*" OR<br>"Oropharyngeal carcinoma" OR<br>"Head and neck cancer" OR<br>"Head and neck tumo*" OR<br>"Head and neck carcinoma" OR<br>"Facial cancer" OR<br>"Facial tumo*" OR<br>"Facial carcinoma" OR<br>"Tongue cancer" OR<br>"Tongue tumo*" OR<br>"Tongue carcinoma" OR<br>Glossectom* OR<br>Post-glossectom* OR<br>Postglossectom*))<br><br>AND<br><br>TS=("movement" OR<br>"articulation" OR<br>"speech" OR<br>"intelligibility" OR<br>"voice" OR<br>"acousti*" OR<br>"phoneti*" OR | TS = abstract, title, and keywords |

| | | "speech perception" OR<br>"Speech therapy" OR<br>"tongue displacement" OR<br>"Tongue motion" OR<br>"Tongue positio*" OR<br>"lingual movement" OR<br>"Lingual displacement" OR<br>"Jaw displacement" OR<br>"Tongue movement" OR<br>"Jaw movement" OR<br>"asymmetr*" OR<br>"symmetr*" OR<br>"Lip displacement" OR<br>"Lip movement" OR<br>"Lip aperture" OR<br>"concav*" OR<br>"tongue tip elevation"))<br><br>AND<br><br>TS=("magnetic resonance imag*" OR<br>"MRI" OR<br>"rt-MRI" OR<br>"rtMRI" OR<br>"Real-time MRI" OR<br>"cine-MRI" OR<br>"ultrasound" OR<br>"UTI" OR<br>"ultrasound tongue imaging" OR<br>"EMA" OR<br>"Electromagnetic articulography" OR<br>"EPG" OR<br>"Electropalatography" OR<br>"Palatography" OR<br>"vocal tract" OR<br>"linguopalatal contact" OR<br>"Videofluoroscop*" OR<br>"X-ray" OR<br>"X-ray microbeam") | |
| Embase | https://www.embase.com/#advancedSearch/default | "mouth tumor"/de OR<br>"oropharynx tumor"/de OR<br>"face tumor"/de OR<br>"tongue tumor"/de OR<br>"head and neck tumor"/de OR<br>"Oral squamous cell carcinoma":ti,ab,kw OR<br>"squamous cell carcinoma":ti,ab,kw OR<br>"Oral cancer":ti,ab,kw OR<br>"Oral tumo*":ti,ab,kw OR<br>"Oral carcinoma":ti,ab,kw OR<br>"Mouth cancer":ti,ab,kw OR<br>"Mouth tumo*":ti,ab,kw OR<br>"Mouth carcinoma":ti,ab,kw OR<br>"Oropharyngeal cancer":ti,ab,kw OR<br>"Oropharyngeal tumo*":ti,ab,kw OR | ti,ab,kw = limit search to title, abstract, and author keywords<br><br>de = index term |

| | | | "Oropharyngeal carcinoma":ti,ab,kw OR<br>"Head and neck cancer":ti,ab,kw OR<br>"Head and neck tumo*":ti,ab,kw OR<br>"Head and neck carcinoma":ti,ab,kw OR<br>"Facial cancer":ti,ab,kw OR<br>"Facial tumo*":ti,ab,kw OR<br>"Facial carcinoma":ti,ab,kw OR<br>"Tongue cancer":ti,ab,kw OR<br>"Tongue tumo*":ti,ab,kw OR<br>"Tongue carcinoma":ti,ab,kw OR<br>Glossectom*:ti,ab,kw OR<br>Post-glossectom*:ti,ab,kw or<br>Postglossectom*:ti,ab,kw<br><br>AND<br><br>"speech disorder"/de OR<br>"speech intelligibility"/de OR<br>movement:ti,ab,kw or<br>articulat*:ti,ab,kw or<br>speech:ti,ab,kw or<br>intelligibility:ti,ab,kw or<br>voice:ti,ab,kw or<br>acousti*:ti,ab,kw or<br>phoneti*:ti,ab,kw or<br>"speech perception":ti,ab,kw or<br>"Speech therapy":ti,ab,kw or<br>"tongue displacement":ti,ab,kw or<br>"Tongue motion":ti,ab,kw or<br>"Tongue positio*":ti,ab,kw or<br>"lingual movement":ti,ab,kw or<br>"Lingual displacement":ti,ab,kw or<br>"Jaw displacement":ti,ab,kw or<br>"Tongue movement":ti,ab,kw or<br>"Jaw movement":ti,ab,kw or<br>"Lip displacement":ti,ab,kw or<br>"Lip movement":ti,ab,kw or<br>"Lip aperture":ti,ab,kw or<br>asymmetr*:ti,ab,kw or<br>symmetr*:ti,ab,kw or<br>concav*:ti,ab,kw or<br>"tongue tip elevation":ti,ab,kw<br><br>AND<br><br>"magnetic resonance imag*":ti,ab,kw or<br>MRI:ti,ab,kw or<br>"rt-MRI":ti,ab,kw or<br>rtMRI:ti,ab,kw or<br>"Real-time MRI":ti,ab,kw or<br>"cine-MRI":ti,ab,kw or<br>ultrasound:ti,ab,kw or<br>UTI:ti,ab,kw or<br>"ultrasound tongue imaging":ti,ab,kw or<br>EMA:ti,ab,kw or | |

| | | | "Electromagnetic articulography":ti,ab,kw or<br>EPG:ti,ab,kw or<br>Electropalatography:ti,ab,kw or<br>Palatography:ti,ab,kw or<br>"vocal tract":ti,ab,kw or<br>"linguopalatal contact":ti,ab,kw or<br>Videofluoroscop*:ti,ab,kw or<br>X-ray:ti,ab,kw or<br>"X-ray microbeam":ti,ab,kw | |

**Appendix B:** Data extraction form

| Category | Variable | Definition | Example Bressman et al. (2007) |
|---|---|---|---|
| **Identification** | Authors | Last names of the study authors | Bressmann, Ackloo, Heng, and Irish |
| | Years of publication | Year the article was published | 2007 |
| | Keywords | Include keywords from journal | Not specified |
| | Full APA reference | Complete reference, including DOI link | Bressmann, T., Ackloo, E., Heng, C. L., & Irish, J. C. (2007). Quantitative three-dimensional ultrasound imaging of partially resected tongues. *Otolaryngology—Head and Neck Surgery, 136*(5), 799-805. https://doi.org/10.1016/j.otohns.2006.11.022 |
| **General study characteristics** | Study design | Choose the study design from the following:<br>(a) Cross-sectional study<br>(b) Case-control study<br>(c) Case report<br>(d) Case series<br>(e) Cohort study<br>(f) Other (specify) | (f) Other (specify)<br>Cohort study with control subjects |
| | Language (article) | Language in which the study was written | English |
| | Location | Country in which the study was conducted | Canada |
| **Participant characteristics*** | Sample size:<br>- Total<br>- OC<br>- HC<br>- Other | Define the:<br>- total sample size of the participant group<br> - sample size of the OC group<br> - sample size of the HC group<br> - sample size of potential other clinical group | Total: 24<br>OC group: 12<br>HC group: 12 |
| | Age:<br>- OC<br>- HC | Define the:<br>- mean age, SD and range of OC participants<br>- mean age, SD and range of HC participants | OC group:<br>M = 45.67<br>SD = 10.39<br>R = 31-58<br><br>HC group:<br>M = 26.0<br>SD = 11.6<br>R = 20-31 |

|  | Gender:<br>- OC<br>- HC | Define the:<br>- gender distribution of OC participants<br>- gender distribution of HC participants | OC group: 8 M 4 F<br><br>HC group: 6 M 6 F |
|---|---|---|---|
|  | Language (study) | Language background of the participants | Canadian English |

* If the article reports several experiments with different participant groups, provide this information for every group. If the article details a longitudinal study in which the participant pool suffered from attrition, the information is provided per point of assessment.

| **Tumour characteristics** | TNM staging | Define the:<br>- T classification<br>- N classification<br>- M classification | Not specified |
|---|---|---|---|
|  | Location | Choose from the following:<br>(a) Anterior tongue (anterior ⅔)<br>(b) Posterior tongue (posterior ⅓)<br>(c) Retromolar trigone<br>(d) Soft palate<br>(e) Buccal mucosa<br>(f) Other (specify) | (a) Anterior tongue (anterior ⅔) 7 patients<br><br>(a) Anterior tongue (anterior ⅔) + (b) Posterior tongue (posterior ⅓): 5 patients |
| **Treatment characteristics** | Treatment modality? | Which treatment modality was used?<br>(a) Surgery<br>(b) Radiation-based | (a) Surgery |
|  | Surgical procedure | What type of procedure was used? | Partial lateral glossectomy |
|  | Reconstruction | Choose from the following:<br>(a) Primary closure<br>(b) Local flap<br>(c) Free flap (e.g., RFFF)<br>(d) Other (specify) | (a) Primary closure: 7 patients<br><br>(c) Free flap: 5 patients |
|  | Radiation type | Specify the type of radiation | NA |
|  | Radiation dosage | How many Gy per fraction? | NA |
| **Experiment information** | Experimental procedure | Detail the experimental procedure (e.g., number of tasks and/or sessions) | There was one session during which one task was administered: producing sustained phonemes while the experimenter made a swipe from the chin to the border of the thyroid cartilage. |

|  | Experiment duration | Define the experiment duration (if provided) | NA |
|---|---|---|---|
|  | Experiment goal | Detail the goal of the experiment | Identify biomechanical commonalities and differences between patients reconstructed with flaps and patients with local defect closures. |
| **Task Information**** | Method | Which method was used (e.g., UTI, real-time-MRI) | 3D UTI |
|  | Type of task | What type of task was used? | Sustained phonemes |
|  | Task procedure | Detail the testing procedure of the task | The subjects sustained the following English phonemes: /a/, i /, /o/, /s/, /S/, /r/, /l/, /n/, /N/ |
|  | Data acquisition | How was the data acquired? | For the data acquisition, subjects were seated upright on an office chair with their head slightly overextended. The transducer was held in a coronal scanning position and swept from the chin to the upper border of the thyroid cartilage.<br><br>As the probe moved during the data acquisition, no holder was used. |
|  |  | Which precautions were taken to ensure high quality data (e.g., was the UTI probe secured?) | NA |
|  | Task duration | If provided, define task duration (in time or number of repetitions/stimuli) | 9 phonemes were repeated 3 times, giving 27 trials. One swipe took 2-3 seconds. |
|  | Task goal | Detail the goal of the task (if different from experiment goal) | NA |
|  | Task measure | Detail the measure of the task | Three sagittal slices (left, midsaggital, and right) were made with five points per slice (two anterior, one central, and two posterior points). This gave 15 data points per tongue scan with their accompanying coordinates. |
|  | Task analysis | Detail the analysis of the task | **Principal component analysis** where components load onto the 15 measurement points denoting the specific regions of the tongue<br><br>**Concavity index** which expresses how convex or concave the tongue is along its whole length<br><br>**Symmetry index** which expresses the symmetry between the two lateral free margins of the tongue |
| ** Each relevant task/session will be described separately if multiple tasks are reported on | | | |
| **Results quantitative*** | Level of significance | What was the threshold of significance? (e.g., 0.05) | 0.05 |
| | Statistical test used | Specify the statistical test used (e.g., ANOVA) | **PCA** with the measurement points of the tongue scans across five groups (controls, local closure pre- and post-op, RFFF pre- and post-op)<br><br>**One-way ANOVA** between concavity / symmetry index and five participant groups (controls, local closure pre- and post-op, RFFF pre- and post-op). Bonferroni post-hoc tests<br><br>**Pearson correlation** between concavity and symmetry index |
| | Descriptive statistics | Provide the mean, SD, and range of task scores for the OC participants and control group | **PCA:**<br>HC:<br>two PCs explained 88.41% of the variance<br>- PC1: 47.38% variance, anterior tongue<br>- PC2: 41.03% variance, posterior tongue<br><br>OC-closure:<br>Pre: two PCs explained 82.43 percent of the variance<br>- PC1: 44.18.% variance, anterior tongue<br>- PC2: 38.25% variance, posterior tongue<br><br>Post: two PCs explained 79.83 percent of the variance<br>- PC1: 48.60% variance, anterior tongue<br>- PC2: 31.23% variance, posterior tongue<br><br>OC-RFFF:<br>Pre: two PCs explained 82.41 percent of the variance<br>- PC1: 46.41% variance, anterior tongue<br>- PC2: 35.99% variance, posterior tongue<br><br>Post: three PCs explained 86.73 percent of the variance<br>- PC1: 46.25% variance, anterior tongue<br>- PC2: 26.24% variance, center posterior tongue<br>- PC3: 14.25% variance, posterior left tongue<br><br>**Concavity:**<br>HC:<br>1.07 cm (SD = 1.56)<br><br>OC-closure:<br>Pre: 1.12 cm (SD = 2.09) |

| | | | Post: 0.91 cm (SD = 0.8) <br><br> <u>OC-RFFF</u>: <br> Pre: 0.5 cm (SD = 1.31) <br> Post: 0.3 cm (SD = 1.73) <br><br> **Symmetry:** <br> <u>HC</u>: <br> 0.86 cm (SD = 0.56) <br><br> <u>OC-closure:</u> <br> Pre: 1.26 cm (SD = 0.91) <br> Post: 1.68 cm (SD = 1.16) <br><br> <u>OC-RFFF</u>: <br> Pre: 1.37 cm (SD = 1.08) <br> Post: 1.9 cm (SD = 1.14) |
|---|---|---|---|
| | Task scores: comparison between groups | If the results of the OC participants were compared to control participants, provide: <br> - results of the comparison <br> - (significant) differences between groups | **Correlation between Concavity and Asymmetry** <br> $r = 0.05$, $p = 0.13$ <br><br> **Correlation between Concavity, Asymmetry and speech acceptability** <br> - Moderate positive correlation between change in concavity and post-operative speech acceptability ($r = 0.69$, $p < 0.02$). <br> - No other significant results <br><br> **Concavity:** <br> - OC-RFFF post-op had lower concavity values as compared to all other groups ($p < 0.05$ in all cases). <br> - No other significant results <br><br> **Asymmetry** <br> - HC had significantly lower asymmetry values as compared to all other groups ($p < 0.001$ in all cases) |
| | Task scores: comparisons within groups | If the results of OC participants were compared within the group, provide: <br> - results of the comparison <br> - (significant) differences between groups | **Concavity** <br> - No significant results <br><br> **Asymmetry** <br> - OC-RFFF had significantly higher asymmetry scores as compared to OC-closure, both pre- and post-op ($p < 0.05$ in all cases) <br> - No other significant results |
| *** If several tasks were used, each task will be reported separately (unless data was pooled) | | | |

| **Results qualitative** | Detail any descriptive results (e.g., error patterns) | NA |
|---|---|---|
| **Individual data** | Is individual data provided for OC participants? Choose from the following:<br>  (a) Yes<br>  (b) No<br><br>If yes, then specify results. | (b) No |